# A femto-Tesla DC SQUID design for quantum-ready readouts

I. Sochnikov[1,2], D. Davino[1], B. Kalisky[3, 4]

[1]Physics Department, University of Connecticut, Storrs, CT USA, 06269

[2]Institue of Material Science, University of Connecticut, Storrs, CT USA, 06269

[3]Department of Physics, Bar-Ilan University, Ramat-Gan, Israel

[4]Institute of Nanotechnology and Advanced Materials, Bar-Ilan University, Ramat-Gan, Israel

## Abstract

Among some of the current uses of the DC Superconducting QUantum Interference Devices (SQUIDs) are qubit-readouts and sensors for probing properties of quantum materials. We present a rather unique gradiometric niobium SQUID design with state-of-the-art sensitivity in the femto-Tesla range which can be easily tuned to specific readout requirements. The sensor is a next generation of the fractional SQUIDs with tightly optimized input coil and a combination of all measures known for restraining parasitic resonances and other detrimental effects. Our design combines the practical usefulness of well-defined pickup loops for superior imaging kernel and tunable-probing applications with the fractionalization approach to reduce undesired inductances. In addition, our modeling predicts small dimensions for these planar sensors. These features make them of high relevance for material studies and for detection of magnetic fields in small volumes, e.g. as part of a cryogenic scanning quantum imaging apparatus for efficient diagnostics and quantum device readouts. This manuscript will benefit scientists and engineers working on quantum computing technologies by clarifying potential general misconceptions about DC SQUID optimization alongside the introduction of the novel flexible compact DC SQUID design.

# Introduction

Superconducting QUantum Interference Devices (SQUIDs) remain one of the least perturbative and most sensitive magnetic field detection technologies available today. The SQUIDs rely on the property of the singly-valued wavefunction along the SQUID circumference, which leads to its periodicity with the flux through the SQUIDs contour (*1*). Nearly any physical property of a SQUID becomes sensitive to flux in a quantized way allowing it to be used as a sensor of magnetic flux or field. SQUIDs range from large devices used in bulk material characterization, living organism signals detection, and geological systems (*1*) to sub-micron size sensors (*2*) at superior signal levels leading to thrilling discoveries in quantum materials (*3–8*). A particularly underexplored area is a utilization of still compact but extremely high field (integrated flux) sensitivity SQUID sensors in a scanning setup. They are ideally suited for challenging applications such as diagnostics of parasitic surface spins on full-scale wafers of materials (e.g. monolayers), and even timelier as non-perturbing qubit-readouts (*9–16*). By using a qubit readout SQUID on a scanning platform (*17*) a *tunable non-perturbative* electromagnetic quantum coupling may be realized which is not possible in readouts fabricated on the same chip as the qubit (Figure 1). Another possible significance of such devices is in quantum-classical interfacing (*18*) where the heat dissipation is removed from a qubit chip by using a remote scanning SQUID, which minimizes backaction and minimizes effects from poisoning phonons and quasiparticles (*19–21*). Thus, oftentimes, SQUIDs are desirable on a millimetric length scale for non-perturbative adjustable coupling and large area or large cross section field integration for efficient signal collection and reduced effects from polluting processes.

In this work we first review the sensitivity concerns in DC SQUIDs and then show modeling and design of practical gradiometric SQUID sensors for the femto-Tesla range (Figure 2, Figure 3, Figure 4) for the scanning imaging and probing applications mentioned above which is a distinct designation from some previous femto-Tesla designs (*22*). Our calculations (Figure 5) predict flux noise figure-of-merit similar to high-coherence qubits and devices (*23–25*) substantiating the prospective effectiveness of these devices (Figure 6) for quantum-ready readouts.

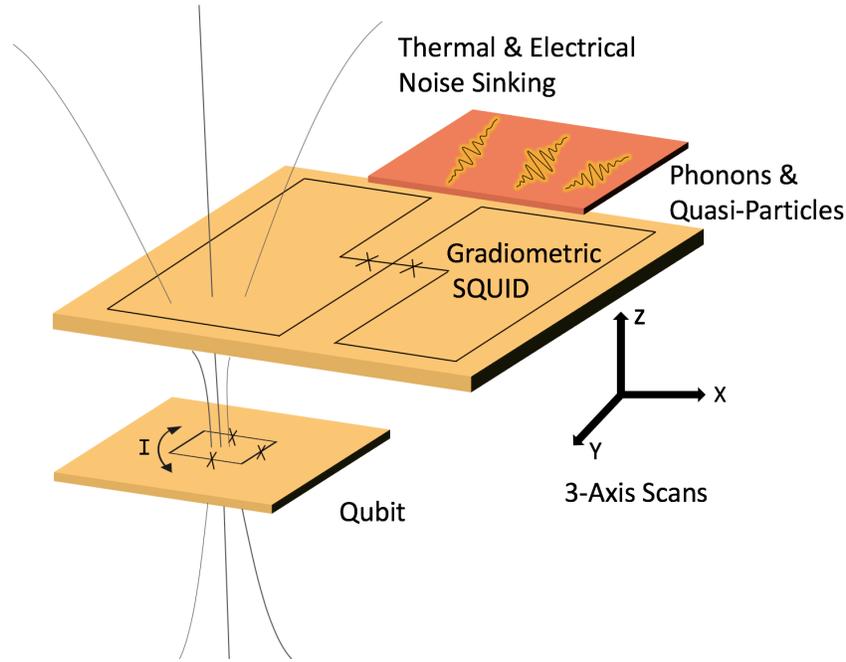

*Figure 1. Scanning SQUID readout of qubits allows for in-situ backaction and coupling tuning and optimization. Detrimental noises from photons, phonons and quasi-particles can be efficiently removed in this proposed setup to enhance qubit coherence (see more details in Appendix).*

## Sensitivity considerations for DC SQUIDs

The theoretical ultimate energy sensitivity of a simple Direct Current SQUID (DC SQUID) was determined to be (*26–28*):

$$\epsilon = \frac{\phi_n^2}{2L} = 16 k_B T \sqrt{LC} \qquad (1)$$

where $L$ is the total inductance of the squid loop and $C$ is the capacitance of each Josephson junction, which yields the flux noise

$$\phi_n = 4\sqrt{2}(k_B T)^{1/2} L^{3/4} C^{1/4} \qquad (2)$$

This value is the approximate theoretical limit expected with available DC SQUID technologies at finite temperatures (at very low temperatures there is another quantum-fluctuations term limiting the possible sensitivity (*29*)). In the current best SQUID systems, the noise level is typically $\sim 10^{-7} - 10^{-6} \Phi_0/\sqrt{Hz}$ (*30*, *31*, *1*, *32–35*, *7*). Achieving even these values is not trivial. This figure of merit is relevant to white noise levels at frequencies >10…100 Hz. Below ~10…100 Hz the so called $1/f$ noise is dominant, but in particularly carefully fabricated devices the $1/f$ component is not tremendously high, making the design considerations in this manuscript valid also for low frequencies with somewhat lesser sensitivity (DC limit). Further progress is limited by Josephson-junction technologies, and also by parasitic noise sources such as charge noise from the dielectrics and from paramagnetic-like spin fluctuations on the surfaces of the metals used in SQUID fabrication (*36*). Thus, a very good level of experimental noise is typically $\sim 10^{-6} \Phi_0/\sqrt{Hz}$. Note that these are white noise figures and low frequency $1/f$ noise is typically worse, however white noise figures are what is typically referred to in literature.

The detected external flux couples to the SQUID either directly to the SQUID loop $L$ or through additional superconducting coils, loops, or transformers (*37*). As the applied flux $\phi$ is changed, the current-voltage characteristics of the SQUID oscillates with a period of $\Phi_0$ (the flux quantum); the critical current modulation depth $\Delta I_M/2I_O$ depends on the important parameter $\beta = 2LI_0/\Phi_0$ holding the information about the modulation depth, where $I_O$ is the zero-flux critical current. For $\beta = 1$, the modulation depth is ~50% or $I_0$ (*30*). Additionally, the current-voltage characteristic is single-valued if another parameter, $\beta_c = 2\pi I_o R^2 C/\Phi_0$, is less than one (*38*), where $R$ is the junction's shunt resistance. These two parameters define the baseline performance of the SQUID. As a practical rule, the values $\beta \sim 1$ and $\beta_c \lesssim 1$ are used to optimize SQUID performance (*26*). Put another way, $I_o R$ should be as large as possible while keeping $\beta_c \lesssim 1$ and $\beta \sim 1$.

The properties above are strongly materials and fabrication process dependent. For the $10 \ \mu A/\mu m^2$ current density in the tri-layer Nb/AlO$_x$/Nb process that we use with SeeQC Inc. (formerly Hypres Inc.), the smallest junction radius in our SQUID is ~0.56 µm, for which we obtain a critical current in the smallest junction of ~10 µA, a capacitance of ~50 $fF$, and a ~5 Ω shunt resistor, while keeping $\beta_c \lesssim 1$. The condition of $\beta \sim 1$ leads to an optimal inductance of the SQUID of $L = 100 \ pF$. In practice, SQUID inductance can be varied by about a factor of 2 without a realistically noticeable compromise in the performance.

On a more practical level, the geometry of the SQUID plays an important role in noise performance of its field-sensitivity (as opposed to flux-sensitivity). This geometry defines the inductances of the SQUID and flux pickup circuits, as well as the parasitic inductances and capacitances which may induce noisy resonances. External and trapped flux rejections are important concerns; flux coupling and flux collection efficiency are also determined by SQUID geometry (*39*). Here, we focused our optimization on these parameters and associated designs to achieve high field-sensitivity, $B_n = \phi_n/A_{eff}$, where $\phi_n$ is the intrinsic flux noise of the SQUID defined above and $A_{eff}$ is the effective area of the SQUID, which generally considers flux focusing and shielding effects (*29, 39–42*). For square washer-like SQUIDs (Figure 2) with $D >> \lambda$ and $w >> \lambda$, where $\lambda$ is the relevant magnetic penetration depth (London or Pearl). The effective area is not the same as the simple geometrical size of the washer and is determined by $A_{eff} \approx D * (D + 2w)$. The inductance of this square washer SQUID is $L = \mu_o (D + w) \left(\frac{2}{\pi}\right) \left(\ln\left(1 + \frac{D}{w}\right) + 0.5\right)$, for D/w≥10, and $L = \mu_o D \left(\frac{2}{\pi}\right) \left(\ln\left(5 + \frac{D}{w}\right) + 0.25\right)$ for D/w≥10, and $L = 1.25 \ \mu_o D$ otherwise (*29*). The washer thickness, $d$, usually enters as a parameter coupled to the penetration depth, the conditions above, $D >> \lambda$ and $w >> \lambda$, mean that we assumed the thickness of the washer $d >> \lambda$. In practice, if we are using a few hundred nanometers thick niobium layers with the London penetration depth of ~80 nm, and lateral sizes larger than 1-2 micrometers, the conditions above are well satisfied. Using these expressions combined with the flux noise expressions above, we calculate the SQUID field sensitivity as shown in Figure 5 (note, the approximate, but practically convenient, expressions used for inductances result in artifacts appearing as kinks in the plots). For a SQUID (single, not gradiometer) pickup area of ~5x5 mm², signal resolution on the order of 5 $fT/\sqrt{Hz}$ is expected, which makes femto-Tesla signals attainable within a few seconds of averaging time. However, 1 $fT/\sqrt{Hz}$ requires a much larger pickup area ($\gtrsim$ 10x10 mm²) in this simple geometry.

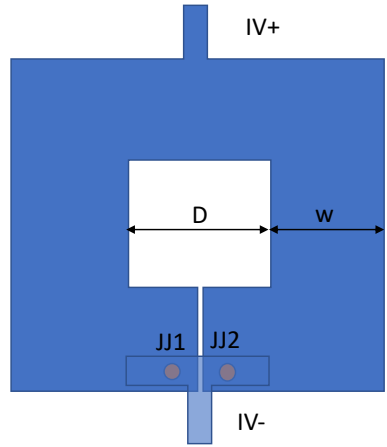

*Figure 2. Square washer SQUID geometry, which we use as a basic element in our calculations and designs. IV+ and IV- are biasing leads. JJ1 and JJ2 are Josephson junctions. The square SQUID provides efficient flux collection on a small chip area as well as efficient coupling to input coils (39).*

One intuitive (but typically incorrect) suggestion for improving the field sensitivity of SQUIDs is to increase the number of turns of the SQUID loop, thus effectively increasing the phase drop (proportional to the flux) on the Josephson junctions. However, the main problem with this solution is that the inductance of the SQUID grows rapidly with the number of turns (usually as the square of the number of turns) and does not outpace the gain from the effective larger pickup (linearly proportional to the number of turns), especially for narrow line loops (*43*). This increase happens because the loops must be large (millimeters) for our target field sensitivities, yielding high single-loop inductances. A better strategy is to reduce the detrimental effects of large inductance rather than only make the SQUID pickup area effectively larger. Note that while this approach may be useful for very small SQUIDs (nano-SQUIDs and micro-SQUIDs) with small inductance, such SQUIDs are not relevant to field-sensitivity devices.

## Flux input configuration

One way to improve the field performance of the SQUID with direct flux input is to connect several SQUID loops of smaller dimensions side-by-side in order to increase the overall field collection area and minimize parasitic capacitive coupling (*44–50*). In this geometry the total inductance of the SQUID is $\sim L/N$, where $L$ is the individual loop inductance and $N$ is the number of loops. However, the improvements in the size of the SQUID are incremental – in the range of a couple of tens of percent reduction (calculated data not shown). This approach usually works well for somewhat less demanding target fields (smaller SQUID pickup area), in this case, the improvement can be more substantial (*44–50*). Besides, the multi-loop SQUID is not suitable for direct field imaging of nearby objects due to complicated imaging kernel of multiple pickup loops. Below we rationalize that a different (new) parallel loops approach can be very useful in SQUIDs with *input coil circuits*, but not in the direct detection form discussed in the preceding paragraphs.

Generally, the main advantage in using the field coil input circuit is that the large inductance of the input coil can be implemented without a substantial degradation of SQUID performance. Magnetic field or flux can be collected by a fully superconducting loop of a desired shape and then magnetically (rather than electrically as in the direct schemes above) coupled to a SQUID (*39*). This approach keeps the SQUID inductance small while the pickup coil inductance can be larger than that achieved via direct schemes without a large penalty in intrinsic SQUID noise. However, measures against the effect of parasitic capacitances and inductances that may lead to undesired resonances must be taken. If not removed properly, these undesired resonances when fed into the SQUID can be amplified by the non-linear current-voltage characteristic of the Josephson junctions, effectively compromising the performance of the SQUID (*51–57*).

The choice of the input coil geometry is not arbitrary. Flux coupling is most efficient when the inductance of the input coil is (ignoring inductances of connecting lines) equal to the inductance of the pickup loop(s): $L_i = L_{pl}$ (if there are two electrically connected pickup loops then coupling is most efficient when $L_i = 2L_{pl}$). In this case, half of the sensed flux (flux through the pickup loop(s)) is screened by the pickup loop portion of the input circuit and the other half by the coil at the input to the SQUID. Thus, the SQUID detects only a fraction of the external flux. The amount of flux detected by the SQUID, $\phi_s$, depends on the mutual inductance between the input coil and the SQUID, $M_i$:

$$\phi_{pl}/\phi_s = 2L_i/M_i \qquad (3)$$

Generally, the mutual inductance is smaller than or equal to the input coil self-inductance. Therefore, the flux noise performance (or energy sensitivity) of such a scheme is worse than that of the direct flux-coupling SQUID described in the beginning of this manuscript. However, a gain in field sensitivity may be obtained for much smaller pickup loop sizes, because the effect on $\beta$ is minimized with the input coil scheme.

In other words, a practical benefit arises because the pickup loop can have a large inductance and a large flux collection area without a large effect on the intrinsic noise: the input coil can be constructed to have a matching inductance, for example by using the Ketchen coupling scheme of a spiral input coil to a wide washer SQUID (*39, 40*). In this case, the self-inductance of the input coil scales with the square of the number of turns $L_i \approx n^2 L$, while the mutual inductance scales as $M_i \approx nL$ (*27, 30, 40*). Thus, the flux noise performance of such a scheme is

$$\phi_{pl} = \frac{2L_i}{M_i} \cdot \phi_s = 2n\phi_s \qquad (4)$$

Combining Eq. (4) with the basic noise performance equation Eq. (2) reveals that the size of the SQUID for sensitivities in the range of 1 fT are substantially smaller than those in the direct coupling schemes. This design therefore constitutes a promising direction to achieve high field-sensitivities. In the next section, we discuss a further improvement with a new scheme of several washers in parallel with input coils connected in series.

## Original design with series input-coils and fractional squid-loops.

The main advantage in using the field coil input circuit is that the large inductance of the input coil can be implemented without a substantial degradation of SQUID performance. However, the flux sensitivity is compromised during matching of the input coil inductance and the pickup loop inductance. This is tolerable for some applications, but often higher sensitivity is needed. Often this compromise is forced by use of a commercial SQUID with a fixed inductance input coil and a customer-provided pickup loop. In other words, when $n$ is large in Eq. (4), it can degrade the performance of a standard input coil with washer setup.

In the following new approach based on the new 'fractional SQUID' designs, we have substantially improved this performance with parallel SQUID loops (Figure 3, Figure 4). Instead of having one washer SQUID loop and an input coil consisting of many turns, we use many parallel washer SQUID loops (*58*), with a single loop input coil each in gradiometric configuration both for the pickup loops and the input washers to cancel the influence of the external field noises when the currents flow with the opposite chirality in the different sub-loops (Figure 3, Figure 4). Each input loop is connected in series, and their parameters and the number of washers, $N$, (and thus the input coil turns) are matched to the pickup loop inductance. When using a single-turn coil, n=1, per each SQUID washer from Eq. (3) and Eq. (4), assuming the same flux through all of the fractional washers, we obtain the total flux sensed by the SQUID as $\phi_s = \frac{M_{1i}}{2L_{1i}} \cdot \phi_{pl}/N$. For washers (*40*, *39*), the mutual inductance is equal to the washer inductance $M_{1i} = L_1$ (Figure 3), and therefore $\phi_{pl} = 2\phi_s$. This is the best flux-at-the-pickup to flux-at-the-SQUID conversion that can be achieved. Our scheme presented here makes the use of this estimate opening the possibility for femto-Tesla sensitivities within relatively small sizes of chips.

It is necessary to have on the order of $N = L_{pl}/L$ SQUID loops connected in parallel, as well as the same number of input coils, leading to a total matching to the pickup loop inductance of tens nH. Figure 5 shows our calculations of these inductances for several target field sensitivities. The calculations are done by looping through $w$ and $D$, calculating the expected SQUID flux noise given the inductances and other parameters of the SQUID and junctions, then translating those to field-noise using the effective area of the pickup loop, and then displaying lines of equal field-noise as a function of the size of the pickup loop. Compellingly, this design promises sub-femto-Tesla sensitivity in about 10x10 mm² footprint per pickup loop. This promises many additional uses in fields ranging from material sciences to neurosciences and in similar or potentially more compact experimental setups (*59–70*).

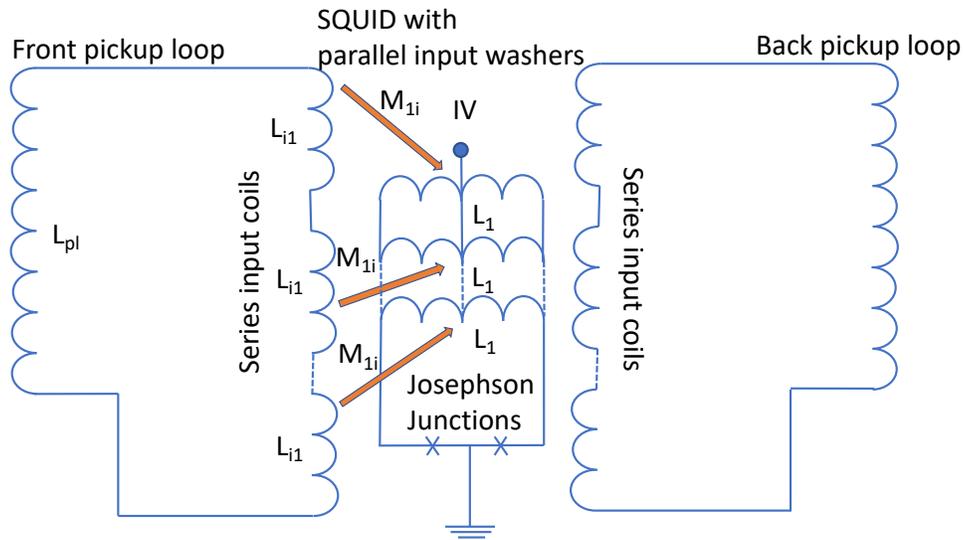

*Figure 3. A simplified circuit diagram of the proposed parallel-washers SQUID gradiometer. The inductance of the N input washers of the SQUID is reduced as $L=L_1/N$ due to the parallel connection to obtain $\beta = 1$. The inductance of the input coils adds in series to $L_i=L_{i1}N$ and is made equal to the pickup loop inductance for the most efficient flux transfer. Left and right pickup circuits couple gradiometrically to the external flux. The design includes only two Josephson junctions, but many loops connected in parallel to those junctions. Very efficient flux transformation can be achieved in this way without compromising the intrinsic noise performance of the SQUID.*

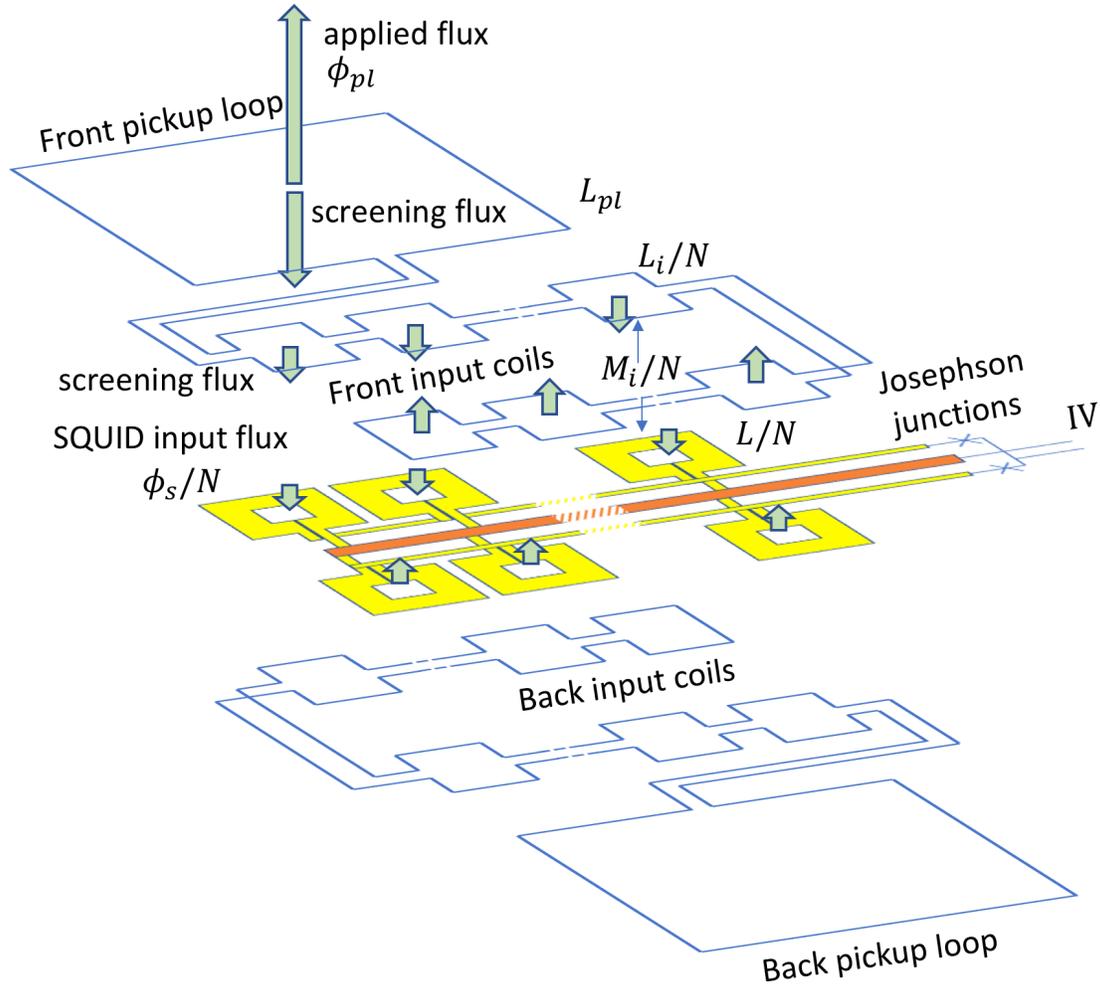

*Figure 4. A wire diagram of our parallel washer gradiometric SQUID (Figure 3) unfolded in a 3-dimensional schematic. The design implements input coils in series showing the coupling between different components of the sensor. There are two gradiometrically configured input circuits with pickup loops and N of one-turn input coils connected in series. The SQUID consists of N parallel washers, configured in a gradiometric fashion, as well. In the actual CAD design, we have further improved this approach by reducing the number of turns in the input coils, reducing parasitic capacitances, and implementing damping resistors to reduce resonances.*

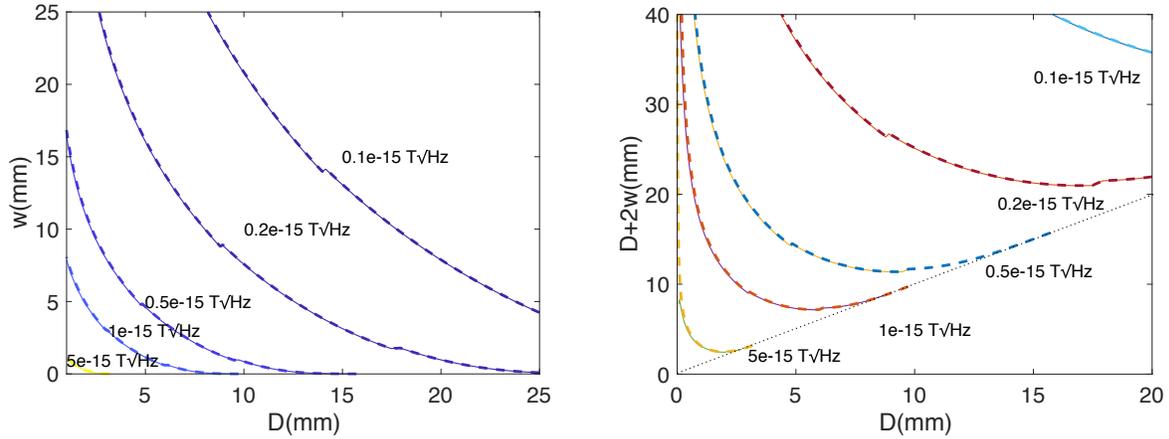

*Figure 5. Sensitivity figures for Ketchen SQUID (solid line) and a fractional-loops washer SQUID (dashed lines) with an input coil. The external flux is collected by a pickup loop and through a series of one-turn input coils coupled to the SQUID. Left: contour plots of the noise in $T/\sqrt{Hz}$ of a SQUID or a pickup loop with an arm width, W, versus the opening size, D. Right: the same sensitivity data plotted as the lateral size of the SQUID versus the SQUID opening size. In bare SQUIDs, to obtain the femto-Tesla sensitivities, the SQUIDs should be quite large, more than 20x20 mm². In the fractional-loops SQUID, and for realistic fabricated chip sizes of ~10x10 mm², we can obtain sensitivities in the range of $sub - fT/\sqrt{Hz}$. Achieving femto-Tesla range sensitivities is projected with the one stage input coil designs – both fractional SQUIDs and Ketchen SQUIDs provide the same expected noise performance.*

## Fabrication-ready gradiometer for the femto-Tesla range

Gradiometric designs (*71*) can offer an external noise rejection of ~10000x or more (*72, 73, 1, 74*). In addition, gradiometers enable more efficient studies of material properties, as they are not susceptible to background signals that do not originate in the samples under study. A field coil can be used to induce local magnetization in the materials. Following the concepts and calculations proposed in the preceding sections, we have designed 50 femto-Tesla gradiometers (and susceptometers) within ~10x20 mm² (Figure 6).

*Table 1. Geometrical parameters of a femto-Tesla SQUID gradiometer.*

| Component | Inner size, D (mm) | Linewidth, w (mm) | Nominal Inductance (nH) | Screened Inductance (nH) |
|---|---|---|---|---|
| Pickup loop | 8.850 | 0.1 | 36.2 | --- |
| One input coil turn | 0.625 | 0.011 | 2.25 | 2.21 |
| One SQUID washer | 0.600 | 0.05 | 1.59 | 1.03 |

Figure 6 depicts the overall layout along with the layers legend. The pickup loops are wide and therefore have moats for flux trapping (*75*) to reduce potential $1/f$ component (so as other components of the SQUID have moats). This design implements 16 parallel washers and the main geometries of the SQUID are summarized in Table 1. Figure A 1 of the appendix contains the details of one realization of the Josephson junctions' region with shunt resistors and an additional damping resistor (*52, 76–79*), which reflect β=1 condition for a 10 µA critical current with 16 parallel washers resulting in ~100 pH SQUID inductance. Figure A 2 of the appendix illustrates a section that includes several parallel washer SQUIDs that are arranged in two parallel columns of 8. These washers have flux trapping moats (*75*). The first 8 and other 8 washers are also gradiometrically configured to reject external noises. The two sets of input coils are electrically isolated, which is also a novel approach. Very small parasitic inductances due to the wide lines and very small capacitances due to very thick oxide (a couple of micrometers of $SiO_2$) are achieved in these designs: the total parasitics are in the range of a few hundreds of pH. The undesired effects of these parasitics should be further minimized by the damping resistors. Figure A 3 shows an individual 1.59 nH washer section, with the flux trapping moats (*75*) and damping resistors emphasized. Overall, these designs show compact devices with greatly enhanced field sensitivity while still being of similar or smaller dimensions than previously (*59–70*). This makes this new design very attractive for practical applications requiring only a few millimeters spatial resolution and yielding an extraordinary field sensitivity in a potentially more compact form-factor than before. Moreover, they can be manufactured at a foundry using conventional superconducting integrated circuits fabrication methods.

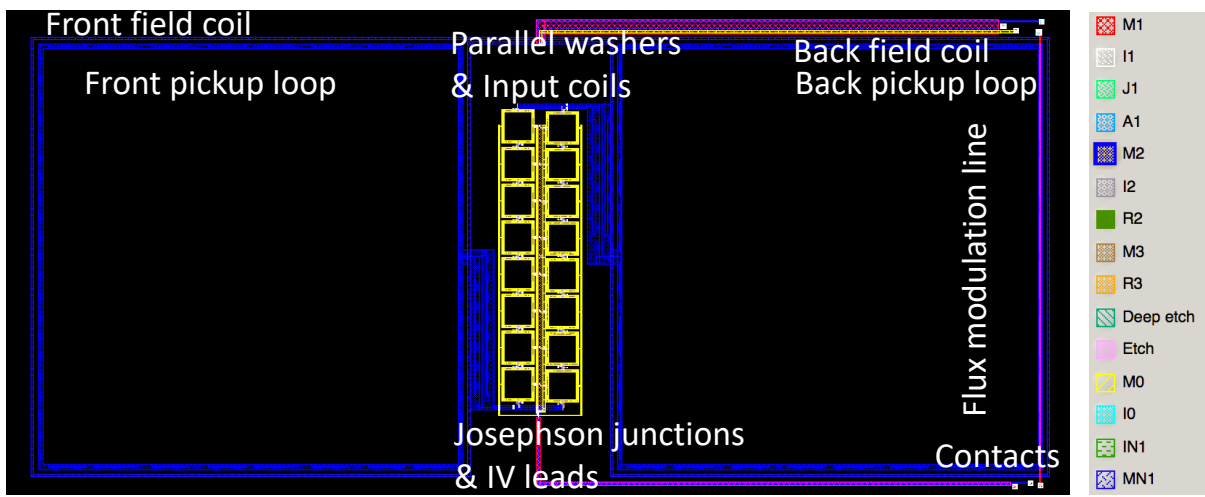

*Figure 6. A representative design of a 100-aT gradiometric magnetometer and susceptometer. This image is a CAD layout with the main components labeled. Additional shield over washers not shown for clarity. The area of the image is 22x10.5 mm$^2$. See Appendix B for more details.*

## Summary

Our detailed optimization methods yielded novel gradiometric SQUIDs designs with projected sensitivities in the femto-Tesla range. Several variants of the designs can be implemented, mainly with varying pickup loop dimensions and numbers of washers. Other parameters that can be varied marginally are the critical currents in Josephson junctions as well as the quantity of such junctions and the values of their shunting resistors. These SQUIDs, even when gradiometric, are of a small size that are practical and relevant for use in various technological and scientific applications, where unprecedented field sensitivity is required. This potentially opens the door to beyond femto-Tesla range, atto-Tesla sensitivity SQUID sensors fabricated as compact chips in the near future, which may open new technological capabilities for discoveries of new phenomena in quantum materials as well as new phenomena in other interdisciplinary fields (*48*, *80–86*) of science and technology, including qubits and quantum-information.

## Acknowledgements

The work by I. S. was in part supported by the US Department of Defense, and the US State of Connecticut. D. D. acknowledges support from the IDEA program at the University of Connecticut.

## Appendices

### Appendix A. Optimizing flux transfer for N fractional loop squid and in-series input coil loops

Here we provide details that lead to the conclusion that the design with fractional loops squid with one-turn in-series input coil loops is optimal for our purposes of detecting efficiently fluxes sensed by large inductance pickup loops but with relatively small dimensions. When a flux $\Delta\phi_{pl}$ from a studied source couples to the pickup loop, it induces a screening current $J_i$ which produces an equal amount of flux distributed in the input transformer as follows (*87*):

$$J_i L_{pl} + J_i L_{i,eff} + J_i L_{i,strip} = \Delta\phi_{pl},$$

where the effective screened inductance of the input coil with $N$ one-turn loops (*1*) is

$$L_{i,eff} \approx (1 - k_i^2 s) L_i = (1 - k_i^2 s) N L_{1i}.$$

The input coil coupling constant is $k_i = \frac{M_1}{\sqrt{L_1 L_{1i}}} = \sqrt{\frac{L_1}{L_{1i}}}$, $s$ is defined in ref. (*1*) and is on the order of 0.04 in our example above, $L_i$ is the total unscreened input coil inductance, $L_{1i}$ is the inductance of the unscreened one-input coil segment out $N$ connected in series, $M_1$ is the mutual inductance between one segment of the input coil and one fractional SQUID washer.

In the classic washer geometry, the inductance of the input coil (an $n$-turn secondary coil) actually coupled to the washer (a 1-turn wide primary coil) is equal to the washer inductance due to an almost perfect imaging of the secondary by the currents in the primary plus the strip inductance of the secondary (*1, 30*):

$$L_{1i} = n^2 L_1 + nL_{1,strip}.$$

The flux transfer factor is the ratio between the externally applied pickup loop flux $\Delta\phi_{pl}$, and the flux sensed by the SQUID, $\Delta\phi_s = M_1/N \cdot NJ_i = k_i\sqrt{L_1 L_{1i}} J_i$ (30, 87):

$$F = \frac{k_i M_1 \sqrt{L_1 L_{1i}}}{L_{pl} + (1 - k_i^2 s)NL_{1i}}$$

By taking the derivative of this expression with respect to the pickup loop inductance, it can be easily shown that this factor is maximized when $L_{pl} = (1 - k_i^2 s)NL_{1i}$ yielding the maximum factor

$$F_{max} = \frac{k_i/N}{2\sqrt{(1-k_i^2 s)}} \sqrt{\frac{L_1}{L_{1i}}},$$

Typically for wide washers and narrow input coils fabricated with standard lithography, $k_i \approx 0.7 - 0.9$ and $s$ is on the level of a few percent (1), therefore

$$F_{max} \approx \frac{1}{2N} \sqrt{\frac{L_1}{L_{1i}}}.$$

Inserting known expressions for the washer and the input coil inductances (1, 88) gives

$$F_{max} \approx \frac{1}{2N} \sqrt{\frac{L_h + L_{sl}}{n^2\left(L_h + \frac{L_{sl}}{3}\right) + nL_{strip}}},$$

$L_h$ is the geometric inductance of the central hole in a washer (primary coil), $L_{sl}$ the slit inductance, and $L_{strip}$ is the stripline inductance of the secondary line (which essentially represents flux leakage). In relatively large washers (tens or hundreds of microns) we can safely assume $L_h \gg L_{sl}$ for finding the optimal $n$. Further, in our case $L_{strip}$ is on the order of 100 pH per one turn of the input coil, while $L_h \approx L = 1.59\ nH$. We can thus write

$$F_{max} \approx \frac{1}{2Nn},$$

which is obviously optimized for $n = 1$ and $L = 1$ as in our design (note, this is not necessarily accurate for small SQUIDs, for example, with a larger fraction of flux leakage in between the primary and the secondary and in the washer slit).

This result echoes the textbook expressions for the classic input coil design for a single washer (72, 87), with the exception of the fact that our expression contains variables that represent a single segment in the input coil and a single turn in the fractional washer. Thus, in our case the Josephson junctions will see a much smaller $\frac{L_1}{N}\left(1 - s_{in}k_i \frac{NL_{1i}}{NL_{1i}+L_{pul}}\right)$ inductance, $s_{in} = 0.5$ in our case as defined in ref. (1), which is beneficial for maintaining a relatively high overall energy sensitivity of the SQUID (meeting the optimization condition for $\beta = 1$).

Note also, that the energy sensitivity of our design is not better than that of the previous approaches and is still fundamentally limited by the same considerations as in the classic cases (*30*, *51*, *87*), but this design is optimized for the compactness of the pickup loop while maintaining high integrated flux (field) sensitivity without compromising the squid inductance. There is one instructive way to understand our results, that the flux from the pickup loop is effectively squeezed by a factor of about 100 (in our example) into a smaller area while still being effectively coupled to a small inductance SQUID. This effect may have further fundamental applications in quantum sensing in general.

## Appendix B. Details of the design layout

Below are enlarged regions of the design presented in Figure 6. They serve as an example of the design details to guide engineering efforts based on our paper.

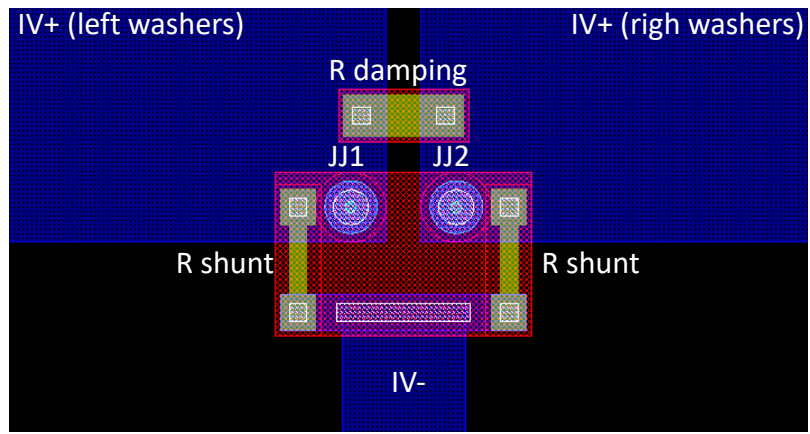

*Figure A 1. Josephson junctions (JJ1 and JJ2). The area of the image is 45x30 $\mu m^2$. This design provides small parasitic inductance and resonance damping.*

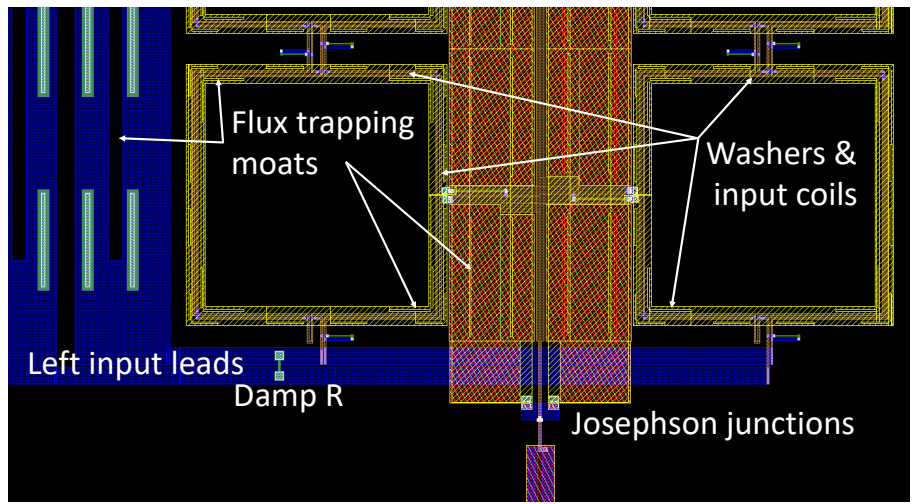

*Figure A 2. Input coil and parallel washers. The area of the image is 2x1 $mm^2$. This is a unique new approach with a single turn input coils connected in series while the washers are connected in parallel to reduce the SQUID inductance.*

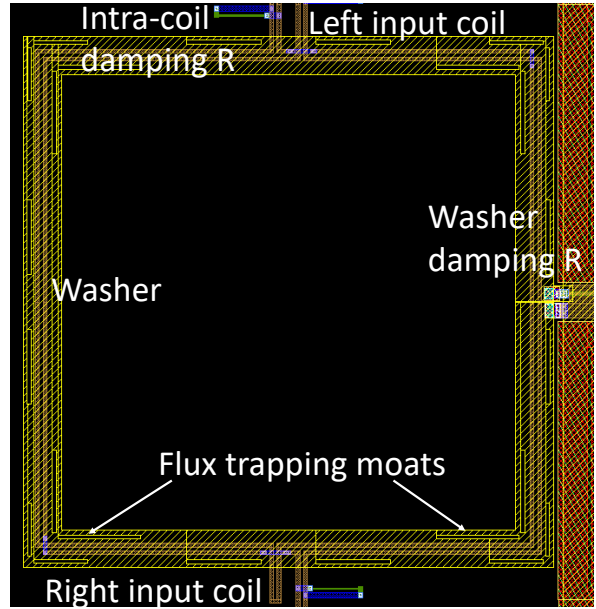

*Figure A 3. Details of the washer section. One-turn loops are used as input coils: one coil for each of the gradiometer pickup loops. Flux traps are implemented as moats in wide Nb sections. Damping resistors are implemented.*

Appendix C. Additional motivations and considerations for scanned SQUID readout and other uses

If one would like to explore experimentally how coupling between a qubit and a readout squid influences the performance of the qubit, a SQUID, such as the one presented here, installed in a scanning apparatus will allow to scan over a large area of a wafer containing many quantum chips and thus effectively replace hundreds if not more on-chip fabricated readout or diagnostics SQUID devices. In a scanning mode, different SQUID-qubit geometrical arrangements (couplings) can be explored more easily than with on-chip readouts.

When a weak coupling is desired, one can think of a (flux) qubit as being a magnetic dipole-like source, where in the weak coupling one would want to position the SQUID far enough away from the qubit. At the same time the collected signal is lost, so to compensate for that, one would want to increase the detection area of the SQUID (to collect more flux). The back action of the noise from the large SQUID on the qubit is smaller than that of a small noisy SQUID well-coupled to a small qubit SQUID, as the noise likely gets dispersed in many directions not just directly in the qubit. Thus, the hypothesis is that, a large fraction of the noise electromagnetic radiation will disperse in the open space (into a cavity and grounded cryostat parts), while a large fraction of the qubit signal can be collected by the large SQUID. More thorough calculations of backactions and experiments will be required to support this hypothesis in the future but these are beyond the scope of this work. In addition, by implementing a remote SQUID, coupling of noise through substrate to qubit is eliminated, again, providing advantages to quantum diagnostics tools.

The kind of SQUIDs presented here could be useful for wafer-scale testing apparatuses for quantum information technologies, like the one introduced recently by Bluefors-Intel-Afore tagged The Cryogenic Wafer Prober for millikelvin range and is said to dramatically help at speeding up the developments of cryogenic quantum devices (*89*). Our sensors can enhance capabilities of such diagnostics tools.

Another possibility of use of these sensors is to test complex noisy quantum systems and to troubleshoot their performance. One such example is when targeting on the individual qubit level may not be useful due to the overall complexity of the tested chip, but the integrated sensitivity to noise and fluctuations is desired. In this case, the spatial resolution is compromised, but finer integrated field resolution could be obtained, allowing for *spectroscopic* studies of complex quantum systems.

# References

1. J. Clarke, A. I. Braginski, *The SQUID Handbook: Fundamentals and Technology of SQUIDs and SQUID Systems* (Wiley-VCH, Weinheim, 1 edition., 2004).

2. José Martínez-Pérez Maria, Koelle Dieter, NanoSQUIDs: Basics & recent advances. *Phys. Sci. Rev.* **2**, 1–27 (2017).

3. I. Sochnikov, A. J. Bestwick, J. R. Williams, T. M. Lippman, I. R. Fisher, D. Goldhaber-Gordon, J. R. Kirtley, K. A. Moler, Direct Measurement of Current-Phase Relations in Superconductor/Topological Insulator/Superconductor Junctions. *Nano Lett.* **13**, 3086–3092 (2013).

4. I. Sochnikov, L. Maier, C. A. Watson, J. R. Kirtley, C. Gould, G. Tkachov, E. M. Hankiewicz, C. Brüne, H. Buhmann, L. W. Molenkamp, K. A. Moler, Nonsinusoidal Current-Phase Relationship in Josephson Junctions from the 3D Topological Insulator HgTe. *Phys. Rev. Lett.* **114**, 066801 (2015).

5. Y. Frenkel, N. Haham, Y. Shperber, C. Bell, Y. Xie, Z. Chen, Y. Hikita, H. Y. Hwang, E. K. H. Salje, B. Kalisky, Imaging and tuning polarity at $SrTiO_3$ domain walls. *Nat. Mat*. **16**, 1203 (2017).

6. D. V. Christensen, Y. Frenkel, Y. Z. Chen, Y. W. Xie, Z. Y. Chen, Y. Hikita, A. Smith, L. Klein, H. Y. Hwang, N. Pryds, B. Kalisky, Strain-tunable magnetism at oxide domain walls. *Nat. Phys.* (2018).

7. K. Bagani, J. Sarkar, A. Uri, M. L. Rappaport, M. E. Huber, E. Zeldov, Y. Myasoedov, Sputtered $Mo_{66}Re_{34}$ SQUID-on-Tip for High-Field Magnetic and Thermal Nanoimaging. *Phys. Rev. Appl.* **12**, 044062 (2019).

8. J. A. Bert, K. C. Nowack, B. Kalisky, H. Noad, J. R. Kirtley, C. Bell, H. K. Sato, M. Hosoda, Y. Hikita, H. Y. Hwang, K. A. Moler, Gate-tuned superfluid density at the superconducting LaAlO$_3$/SrTiO$_3$ interface. *Phys Rev B*. **86**, 060503 (2012).

9. R. H. Koch, G. A. Keefe, F. P. Milliken, J. R. Rozen, C. C. Tsuei, J. R. Kirtley, D. P. DiVincenzo, Experimental Demonstration of an Oscillator Stabilized Josephson Flux Qubit. *Phys. Rev. Lett.* **96**, 127001 (2006).

10. S. Saito, T. Meno, M. Ueda, H. Tanaka, K. Semba, H. Takayanagi, Parametric Control of a Superconducting Flux Qubit. *Phys. Rev. Lett.* **96**, 107001 (2006).

11. M. G. Castellano, F. Chiarello, P. Carelli, C. Cosmelli, F. Mattioli, G. Torrioli, Deep-well ultrafast manipulation of a SQUID flux qubit. *New J. Phys.* **12**, 043047 (2010).

12. H. Deng, Y. Wu, Y. Zheng, N. Akhtar, J. Fan, X. Zhu, J. Li, Y. Jin, D. Zheng, Working Point Adjustable DC-SQUID for the Readout of Gap Tunable Flux Qubit. *IEEE Trans. Appl. Supercond.* **25**, 1–4 (2015).

13. X. Y. Jin, S. Gustavsson, J. Bylander, F. Yan, F. Yoshihara, Y. Nakamura, T. P. Orlando, W. D. Oliver, Z-Gate Operation on a Superconducting Flux Qubit via its Readout SQUID. *Phys. Rev. Appl.* **3**, 034004 (2015).

14. C. Eichler, J. R. Petta, Realizing a Circuit Analog of an Optomechanical System with Longitudinally Coupled Superconducting Resonators. *Phys. Rev. Lett.* **120**, 227702 (2018).

15. E. Leonard, M. A. Beck, J. Nelson, B. G. Christensen, T. Thorbeck, C. Howington, A. Opremcak, I. V. Pechenezhskiy, K. Dodge, N. P. Dupuis, M. D. Hutchings, J. Ku, F. Schlenker, J. Suttle, C. Wilen, S. Zhu, M. G. Vavilov, B. L. T. Plourde, R. McDermott, Digital Coherent Control of a Superconducting Qubit. *Phys. Rev. Appl.* **11**, 014009 (2019).

16. X. Wang, A. Miranowicz, F. Nori, Ideal Quantum Nondemolition Readout of a Flux Qubit without Purcell Limitations. *Phys. Rev. Appl.* **12**, 064037 (2019).

17. J. R. Kirtley, Fundamental studies of superconductors using scanning magnetic imaging. *Rep. Prog. Phys.* **73**, 126501 (2010).

18. R. McDermott, M. G. Vavilov, B. L. T. Plourde, F. K. Wilhelm, P. J. Liebermann, O. A. Mukhanov, T. A. Ohki, Quantum–classical interface based on single flux quantum digital logic. *Quantum Sci. Technol.* **3**, 024004 (2018).

19. Note, normally on-chip DC SQUID can dampen coherence of a qubit substantially, e. g. due to resistive shunts. Using remote SQUID off a qubit chip and at the same time increasing the field integration (pickup) area may reduce such decoherence effects, as the mutual inductance decreases, while the effective signal-to-noise may be increased with the SQUID pickup area.

20. J. Aumentado, M. W. Keller, J. M. Martinis, M. H. Devoret, Nonequilibrium Quasiparticles and 2e Periodicity in Single-Cooper-Pair Transistors. *Phys. Rev. Lett.* **92**, 066802 (2004).

21. U. Patel, I. V. Pechenezhskiy, B. L. T. Plourde, M. G. Vavilov, R. McDermott, Phonon-mediated quasiparticle poisoning of superconducting microwave resonators. *Phys. Rev. B*. **96**, 220501 (2017).

22. D. Drung, H. Koch, An integrated DC SQUID magnetometer with variable additional positive feedback. *Supercond. Sci. Technol.* **7**, 242–245 (1994).

23. P. Kumar, S. Sendelbach, M. A. Beck, J. W. Freeland, Z. Wang, H. Wang, C. C. Yu, R. Q. Wu, D. P. Pappas, R. McDermott, Origin and Reduction of 1/f Magnetic Flux Noise in Superconducting Devices. *Phys. Rev. Appl.* **6**, 041001 (2016).

24. F. Yan, S. Gustavsson, A. Kamal, J. Birenbaum, A. P. Sears, D. Hover, T. J. Gudmundsen, D. Rosenberg, G. Samach, S. Weber, J. L. Yoder, T. P. Orlando, J. Clarke, A. J. Kerman, W. D. Oliver, The flux qubit revisited to enhance coherence and reproducibility. *Nat. Commun.* **7**, 12964 (2016).

25. L. B. Nguyen, Y.-H. Lin, A. Somoroff, R. Mencia, N. Grabon, V. E. Manucharyan, High-Coherence Fluxonium Qubit. *Phys. Rev. X*. **9**, 041041 (2019).

26. C. D. Tesche, J. Clarke, dc SQUID: Noise and optimization. *J. Low Temp. Phys.* **29**, 301–331 (1977).

27. M. B. Ketchen, D. D. Awschalom, W. J. Gallagher, A. W. Kleinsasser, R. L. Sandstrom, J. R. Rozen, B. Bumble, Design, fabrication, and performance of integrated miniature SQUID susceptometers. *IEEE Trans. Magn.* **25**, 1212–1215 (1989).

28. D. Drung, Introduction to Nb-based SQUID Sensors (2016), (available at https://snf.ieeecsc.org/abstracts/cr70-introduction-nb-based-squid-sensors).

29. H. Weinstock, Ed., *SQUID Sensors: Fundamentals, Fabrication and Applications* (Springer Netherlands, 1996), *Nato Science Series E:*

30. M. Ketchen, DC SQUIDs 1980: The state of the art. *IEEE Trans. Magn.* **17**, 387–394 (1981).

31. P. Carelli, M. G. Castellano, High-sensitivity DC-SQUID measurements. *Phys. B Condens. Matter*. **280**, 537–539 (2000).

32. F. Giazotto, J. T. Peltonen, M. Meschke, J. P. Pekola, Superconducting quantum interference proximity transistor. *Nat. Phys.* **6**, 254–259 (2010).

33. A. Ronzani, C. Altimiras, F. Giazotto, Highly Sensitive Superconducting Quantum-Interference Proximity Transistor. *Phys. Rev. Appl.* **2**, 024005 (2014).

34. J. R. Kirtley, L. Paulius, A. J. Rosenberg, J. C. Palmstrom, C. M. Holland, E. M. Spanton, D. Schiessl, C. L. Jermain, J. Gibbons, Y.-K.-K. Fung, M. E. Huber, D. C. Ralph, M. B. Ketchen, G. W. Gibson, K. A. Moler, Scanning SQUID susceptometers with sub-micron spatial resolution. *Rev. Sci. Instrum.* **87**, 093702 (2016).

35. R. N. Jabdaraghi, D. S. Golubev, J. P. Pekola, J. T. Peltonen, Noise of a superconducting magnetic flux sensor based on a proximity Josephson junction. *Sci. Rep.* **7**, 8011 (2017).

36. S. M. Anton, J. S. Birenbaum, S. R. O'Kelley, V. Bolkhovsky, D. A. Braje, G. Fitch, M. Neeley, G. C. Hilton, H.-M. Cho, K. D. Irwin, F. C. Wellstood, W. D. Oliver, A. Shnirman, J. Clarke, Magnetic Flux Noise in dc SQUIDs: Temperature and Geometry Dependence. *Phys. Rev. Lett.* **110**, 147002 (2013).

37. J. A. B. Mates, K. D. Irwin, L. R. Vale, G. C. Hilton, H. M. Cho, An Efficient Superconducting Transformer Design for SQUID Magnetometry. *J. Low Temp. Phys.* **176**, 483–489 (2014).

38. H. Seppa, T. Ryhanen, R. Imoniemi, J. Knuutila, in *Superconducting Technology: 10 Case Studies* (World Scientific, 1991), pp. 1–29.

39. M. B. Ketchen, J. M. Jaycox, Ultra-low-noise tunnel junction dc SQUID with a tightly coupled planar input coil. *Appl. Phys. Lett.* **40**, 736–738 (1982).

40. J. Jaycox, M. Ketchen, Planar coupling scheme for ultra low noise DC SQUIDs. *IEEE Trans. Magn.* **17**, 400–403 (1981).

41. T. Van Duzer, C. Terner, *Principles of Superconductive Devices and Circuits* (Prentice-Hall PTR, Upper Saddle River, N.J, 2nd Edition edition., 1999).

42. E. H. Brandt, Thin superconductors and SQUIDs in perpendicular magnetic field. *Phys. Rev. B*. **72**, 024529 (2005).

43. M. B. Ketchen, K. G. Stawiasz, D. J. Pearson, T. A. Brunner, C. -K. Hu, M. A. Jaso, M. P. Manny, A. A. Parsons, K. J. Stein, Sub-μm linewidth input coils for low Tc integrated thin-film dc superconducting quantum interference devices. *Appl. Phys. Lett.* **61**, 336–338 (1992).

44. P. Carelli, V. Foglietti, Improved multi-loop DC SQUID. *IEEE Trans. Magn.* **19**, 299–302 (1983).

45. S. Zarembiński, T. Claeson, Design of multiloop input circuits for high-$T_c$ superconducting quantum interference magnetometers. *J. Appl. Phys.* **72**, 1918–1935 (1992).

46. D. Drung, S. Knappe, H. Koch, Theory for the multiloop dc superconducting quantum interference device magnetometer and experimental verification. *J. Appl. Phys.* **77**, 4088–4098 (1995).

47. M. Kiviranta, L. Gronberg, J. Hassel, A Multiloop SQUID and a SQUID Array With 1-μm and Submicrometer Input Coils. *IEEE Trans. Appl. Supercond.* **22**, 1600105–1600105 (2012).

48. J. Luomahaara, A. Kemppinen, P. Helistö, J. Hassel, Characterization of SQUID-Based Null Detector for a Quantum Metrology Triangle Experiment. *IEEE Trans. Appl. Supercond.* **23**, 1601705–1601705 (2013).

49. J. Luomahaara, M. Kiviranta, L. Grönberg, K. C. J. Zevenhoven, P. Laine, Unshielded SQUID Sensors for Ultra-Low-Field Magnetic Resonance Imaging. *IEEE Trans. Appl. Supercond.* **28**, 1–4 (2018).

50. M. Kiviranta, L. Grönberg, J. van der Kuur, Two SQUID amplifiers intended to alleviate the summing node inductance problem in multiplexed arrays of Transition Edge Sensors. *arXiv:1810.09122* (2018) (available at http://arxiv.org/abs/1810.09122).

51. J. Knuutila, M. Kajola, H. Seppä, R. Mutikainen, J. Salmi, Design, optimization, and construction of a dc SQUID with complete flux transformer circuits. *J. Low Temp. Phys.* **71**, 369–392 (1988).

52. V. Foglietti, W. J. Gallagher, M. B. Ketchen, A. W. Kleinsasser, R. H. Koch, R. L. Sandstrom, Performance of dc SQUIDs with resistively shunted inductance. *Appl. Phys. Lett.* **55**, 1451–1453 (1989).

53. K. Enpuku, K. Yoshida, Modeling the dc superconducting quantum interference device coupled to the multiturn input coil. *J. Appl. Phys.* **69**, 7295–7300 (1991).

54. T. Ryhanen, R. Cantor, D. Drung, H. Koch, H. Seppa, Effect of parasitic capacitance on DC SQUID performance. *IEEE Trans. Magn.* **27**, 3013–3016 (1991).

55. V. Foglietti, M. E. Giannini, G. Petrocco, A double DC-SQUID device for flux locked loop operation. *IEEE Trans. Magn.* **27**, 2989–2992 (1991).

56. R. Cantor, K. Enpuku, T. Ryhanen, H. Seppa, A high performance integrated DC SQUID magnetometer. *IEEE Trans. Appl. Supercond.* **3**, 1800–1803 (1993).

57. M. E. Huber, P. A. Neil, R. G. Benson, D. A. Burns, A. M. Corey, C. S. Flynn, Y. Kitaygorodskaya, O. Massihzadeh, J. M. Martinis, G. C. Hilton, DC SQUID series array amplifiers with 120 MHz bandwidth. *IEEE Trans. Appl. Supercond.* **11**, 1251–1256 (2001).

58. P. Carelli, M. G. Castellano, G. Torrioli, R. Leoni, Low noise multiwasher superconducting interferometer. *Appl. Phys. Lett.* **72**, 115–117 (1998).

59. R. Ilmoniemi, J. Knuutila, T. Ryhänen, H. Seppä, in *Progress in Low Temperature Physics*, D. F. Brewer, Ed. (Elsevier, 1989), vol. 12, pp. 271–339.

60. W. Vodel, K. Makiniemi, An ultra low noise DC SQUID system for biomagnetic research. *Meas. Sci. Technol.* **3**, 1155–1160 (1992).

61. D. Drung, H. Koch, An electronic second-order gradiometer for biomagnetic applications in clinical shielded rooms. *IEEE Trans. Appl. Supercond.* **3**, 2594–2597 (1993).

62. C. Granata, A. Vettoliere, M. Russo, Miniaturized superconducting quantum interference magnetometers for high sensitivity applications. *Appl. Phys. Lett.* **91**, 122509 (2007).

63. C. Granata, A. Vettoliere, S. Rombetto, C. Nappi, M. Russo, Performances of compact integrated superconducting magnetometers for biomagnetic imaging. *J. Appl. Phys.* **104**, 073905 (2008).

64. M. I. Faley, U. Poppe, K. Urban, R. L. Fagaly, Noise analysis of DC SQUIDs with damped superconducting flux transformers. *J. Phys. Conf. Ser.* **234**, 042009 (2010).

65. A. Vettoliere, C. Granata, S. Rombetto, M. Russo, Modeled Performance of a Long Baseline Planar SQUID Gradiometer for Biomagnetism. *IEEE Trans. Appl. Supercond.* **21**, 383–386 (2011).

66. Y. Zhang, C. Liu, M. Schmelz, H.-J. Krause, A. I. Braginski, R. Stolz, X. Xie, H.-G. Meyer, A. Offenhäusser, M. Jiang, Planar SQUID magnetometer integrated with bootstrap circuitry under different bias modes. *Supercond. Sci. Technol.* **25**, 125007 (2012).

67. A. Tsukamoto, S. Adachi, Y. Oshikubo, K. Tanabe, Design and Fabrication of Directly-Coupled HTS-SQUID Magnetometer With a Multi-Turn Input Coil. *IEEE Trans. Appl. Supercond.* **23**, 1600304–1600304 (2013).

68. K. Yang, H. Chen, X. Kong, L. Lu, M. Li, R. Yang, X. Xie, Weakly Damped SQUID Gradiometer With Low Crosstalk for Magnetocardiography Measurement. *IEEE Trans. Appl. Supercond.* **26**, 1–5 (2016).

69. M. Schmelz, V. Zakosarenko, A. Chwala, T. Schönau, R. Stolz, S. Anders, S. Linzen, H. -. Meyer, Thin-Film-Based Ultralow Noise SQUID Magnetometer. *IEEE Trans. Appl. Supercond.* **26**, 1–5 (2016).

70. J.-H. Storm, P. Hömmen, D. Drung, R. Körber, An ultra-sensitive and wideband magnetometer based on a superconducting quantum interference device. *Appl. Phys. Lett.* **110**, 072603 (2017).

71. C. Granata, A. Vettoliere, Nano Superconducting Quantum Interference device: a powerful tool for nanoscale investigations. *Phys. Rep.* **614**, 1–69 (2016).

72. M. B. Ketchen, Design of improved integrated thin-film planar dc SQUID gradiometers. *J. Appl. Phys.* **58**, 4322–4325 (1985).

73. A. Kittel, K. A. Kouznetsov, R. McDermott, B. Oh, J. Clarke, High $T_c$ superconducting second-order gradiometer. *Appl. Phys. Lett.* **73**, 2197–2199 (1998).

74. A. Chwala, J. Kingman, R. Stolz, M. Schmelz, V. Zakosarenko, S. Linzen, F. Bauer, M. Starkloff, M. Meyer, H.-G. Meyer, Noise characterization of highly sensitive SQUID magnetometer systems in unshielded environments. *Supercond. Sci. Technol.* **26**, 035017 (2013).

75. M. Jeffery, T. Van Duzer, J. R. Kirtley, M. B. Ketchen, Magnetic imaging of moat-guarded superconducting electronic circuits. *Appl. Phys. Lett.* **67**, 1769–1771 (1995).

76. K. Enpuku, T. Muta, K. Yoshida, F. Irie, Noise characteristics of a dc SQUID with a resistively shunted inductance. *J. Appl. Phys.* **58**, 1916–1923 (1985).

77. K. Enpuku, K. Yoshida, S. Kohjiro, Noise characteristics of a dc SQUID with a resistively shunted inductance. II. Optimum damping. *J. Appl. Phys.* **60**, 4218–4223 (1986).

78. K. Enpuku, H. Koch, Washer Resonances of DC Superconducting Quantum Interference Device Coupled to Multiturn Input Coil. *Jpn. J. Appl. Phys.* **32**, 3811–3816 (1993).

79. N. Shimizu, T. Morooka, K. Enpuku, Suppression of Washer Resonance of DC Superconducting Quantum Interference Device by Using New Washer with Additional Slit. *Jpn. J. Appl. Phys.* **33**, L1215–L1217 (1994).

80. J. Flokstra, H. J. M. Ter Brake, E. P. Houwman, D. Veldhuis, W. Jaszczuk, M. Caspari, H. Rogalla, A. Martı́nez, C. Rillo, A 19-channel d.c. SQUID magnetometer system for brain research. *Sens. Actuators Phys.* **27**, 781–785 (1991).

81. K. Isawa, S. Nakayama, M. Ikeda, S. Takagi, S. Tosaka, N. Kasai, Robotic 3D SQUID imaging system for practical nondestructive evaluation applications. *Phys. C Supercond.* **432**, 182–192 (2005).

82. C. Granata, A. Vettoliere, M. Russo, An ultralow noise current amplifier based on superconducting quantum interference device for high sensitivity applications. *Rev. Sci. Instrum.* **82**, 013901 (2011).

83. A. Chwala, J. P. Smit, R. Stolz, V. Zakosarenko, M. Schmelz, L. Fritzsch, F. Bauer, M. Starkloff, H.-G. Meyer, Low temperature SQUID magnetometer systems for geophysical exploration with transient electromagnetics. *Supercond. Sci. Technol.* **24**, 125006 (2011).

84. V. Zakosarenko, M. Schmelz, R. Stolz, T. Schönau, L. Fritzsch, S. Anders, H.-G. Meyer, Femtoammeter on the base of SQUID with thin-film flux transformer. *Supercond. Sci. Technol.* **25**, 095014 (2012).

85. M. I. Faley, J. Dammers, Y. V. Maslennikov, J. F. Schneiderman, D. Winkler, V. P. Koshelets, N. J. Shah, R. E. Dunin-Borkowski, High-T$_c$ SQUID biomagnetometers. *Supercond. Sci. Technol.* **30**, 083001 (2017).

86. C. Granata, A. Vettoliere, O. Talamo, P. Silvestrini, R. Rucco, P. P. Sorrentino, F. Jacini, F. Baselice, M. Liparoti, A. Lardone, G. Sorrentino, in *Sensors*, B. Andò, F. Baldini, C. Di Natale, V. Ferrari, V. Marletta, G. Marrazza, V. Militello, G. Miolo, M. Rossi, L. Scalise, P. Siciliano, Eds. (Springer International Publishing, 2019), pp. 203–209.

87. J. O. Lekkala, J. A. V. Malmivuo, Optimization of a squid vector gradiometer. *Cryogenics*. **25**, 291–303 (1985).

88. M. B. Ketchen, Design considerations for DC SQUIDs fabricated in deep sub-micron technology. *IEEE Trans. Magn.* **27**, 2916–2919 (1991).

89. Intel drives development of quantum cryoprober with Bluefors and Afore to accelerate quantum computing. *Intel Newsroom* (2019).